\documentclass[DIV=calc,paper=a4,fontsize=11pt,twocolumn]{scrartcl} 

\usepackage[english]{babel}
\usepackage[protrusion=true,expansion=true]{microtype}
\usepackage{amsmath,amsfonts,amsthm}
\usepackage[final]{graphicx}
\usepackage{xcolor}
\usepackage[normal,small,hypcap,up,labelfont=bf,textfont=it]{caption}
\usepackage{epstopdf}
\usepackage{subfig}
\usepackage{booktabs}
\usepackage{fix-cm}
\usepackage{amssymb,amsfonts}
\usepackage{dsfont}
\usepackage{bbm}
\usepackage{pstricks}
\usepackage{cite}
\usepackage[utf8]{inputenc}
\usepackage[perpage,symbol*]{footmisc}
\usepackage[varg]{txfonts}
\usepackage{balance}
\usepackage{fancyhdr}
\PassOptionsToPackage{hyphens}{url}\usepackage[pdfencoding=auto,psdextra]{hyperref}
\usepackage{bookmark}
\usepackage{verbatim}
\usepackage{fontenc}
\usepackage{cuted}
\usepackage{enumerate}

\usepackage{bm}
\usepackage{mathrsfs}

\theoremstyle{definition}

\theoremstyle{plain}

\DeclareCaptionFont{mycolor}{\color[HTML]{0000FF}}
\usepackage{sectsty}													
\allsectionsfont{
\color[HTML]{31ADF3}\usefont{OT1}{phv}{b}{n}
}

\sectionfont{
\color[HTML]{31ADF3}\usefont{OT1}{phv}{b}{n}
}

\usepackage{fancyhdr}												
\usepackage{lettrine}
\newcommand{\initial}[1]{%
\lettrine[lines=3,lhang=0.3,nindent=0em]{
\color[HTML]{31ADF3}
{\textsf{#1}}}{}}

\usepackage{titling}															

\newcommand{\HorRule}{\color[HTML]{31ADF3}
\rule{\linewidth}{1pt}%
}

\pretitle{\vspace{-30pt} \begin{flushleft} \HorRule
\fontsize{34}{34} \usefont{OT1}{phv}{b}{n} \color[HTML]{31ADF3} \selectfont
}
\title{Schr\"odinger's Cat:\\Where Does The Entanglement Come From?}					
\posttitle{\par\end{flushleft}\vskip 0.5em}

\preauthor{\begin{flushleft}\large \lineskip 0.5em \usefont{OT1}{phv}{b}{sl} \color[HTML]{31ADF3}}
\author{Radu Ionicioiu\\[8pt]}											
\postauthor{\footnotesize \usefont{OT1}{phv}{m}{sl} \color[HTML]{000000}
Department of Theoretical Physics, Horia Hulubei National Institute of Physics and Nuclear Engineering,\\ Bucharest--M\u{a}gurele, Romania. E-mail: \href{mailto:r.ionicioiu@theory.nipne.ro}{r.ionicioiu@theory.nipne.ro}\\
[10pt]		
\scriptsize\usefont{OT1}{phv}{m}{n} \color[HTML]{31ADF3}{\textbf{Editors: \emph{Jonas Maziero} \& \emph{Danko Georgiev}} }\\[5pt]
\color[HTML]{000000}{Article history: Submitted on September 19, 2017;  Accepted on October 6, 2017; Published on October 7, 2017.}
\par\end{flushleft}\HorRule}

\date{}																				

\newcommand{\bra}[1]{\langle #1 |}
\newcommand{\ket}[1]{| #1 \rangle}

\newcommand{\ignore}[1]{}

\begin{document}
\maketitle
\thispagestyle{fancy} 			
\initial{S}\textbf{chr\"odinger's cat
is one of the most striking paradoxes of quantum mechanics that reveals the counterintuitive aspects of the microscopic world. Here, I discuss the paradox in the framework of quantum information. Using a quantum networks formalism, I analyse the information flow between the atom and the cat. This reveals that the atom and the cat are connected only through a classical information channel: the detector clicks $\rightarrow$ the poison is released $\rightarrow$ the cat is killed. No amount of local operations and classical communication can entangle the atom and the cat, which are initially in a separable state. This casts a new light on the paradox.\\ Quanta 2017; 6: 57--60.}
\begin{figure}[b!]
\rule{245 pt}{0.5 pt}\\[3pt]
\raisebox{-0.2\height}{\includegraphics[width=5mm]{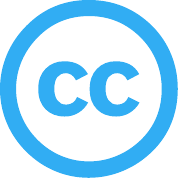}}\raisebox{-0.2\height}{\includegraphics[width=5mm]{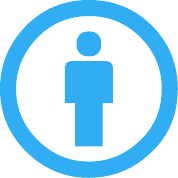}}
\footnotesize{This is an open access article distributed under the terms of the Creative Commons Attribution License \href{http://creativecommons.org/licenses/by/3.0/}{CC-BY-3.0}, which permits unrestricted use, distribution, and reproduction in any medium, provided the original author and source are credited.}
\end{figure}

\section{Introduction}

Schr\"odinger's cat is arguably the most famous and discussed thought experiment in quantum mechanics \cite{Schrodinger1935}. Historically, it is also one of the first to reveal the paradoxical nature of quantum theory when applied to macroscopic objects.
Consider a box containing a cat, a radioactive atom prepared in the excited state~$\ket{0}$ and a detector coupled to the atom. If the atom decays to its ground state~$\ket{1}$, the detector triggers a mechanism which releases a poison and kills the cat. We close the box and we let the isolated system evolve. In the textbook description of the experiment, after some time the state of the atom--cat system is \cite[pp. 373-374]{Peres2002}
\begin{equation}
\ket{\psi_{\textrm f}}= a \ket{0} \ket{alive}+ b \ket{1} \ket{dead}
\label{SC}
\end{equation}
hence the cat is ``simultaneously'' alive and dead, thus the paradox; usually one takes $a=b= \frac{1}{\sqrt 2}$.

The state \eqref{SC} is entangled, as can be seen from the concurrence $C(\ket{\psi_{\textrm f}})= 2|a \cdot b|$; for $|a|= |b|= \frac{1}{\sqrt 2} $ the state is maximally entangled, $C=1$. This brings us to the main question of this article:

\emph{Where does the entanglement come from}?

\section{Contradiction}

Consider the following statements:
\begin{enumerate}[(i)]
\item the initial state of the system is separable, $\ket{0} \ket{alive}$. The cat is in a well-defined state $\ket{alive}$ and is not entangled with the atom;
\item the atom evolves freely to $a\ket{0}+ b\ket{1}$;
\item the only coupling between the atom and the cat is via the detector; there is no {\em entangling quantum interaction} between the two systems;
\item the state \eqref{SC} of the atom--cat system is entangled.
\end{enumerate}
The purpose of this article is to show that there is a contradiction between the first three statements (i)--(iii) and the last one (iv). Briefly, starting with a separable state and having only a classical communication channel between the atom and the cat one cannot generate the entangled state \eqref{SC}.

\begin{figure}[t!]
\centering
\includegraphics[width=85mm]{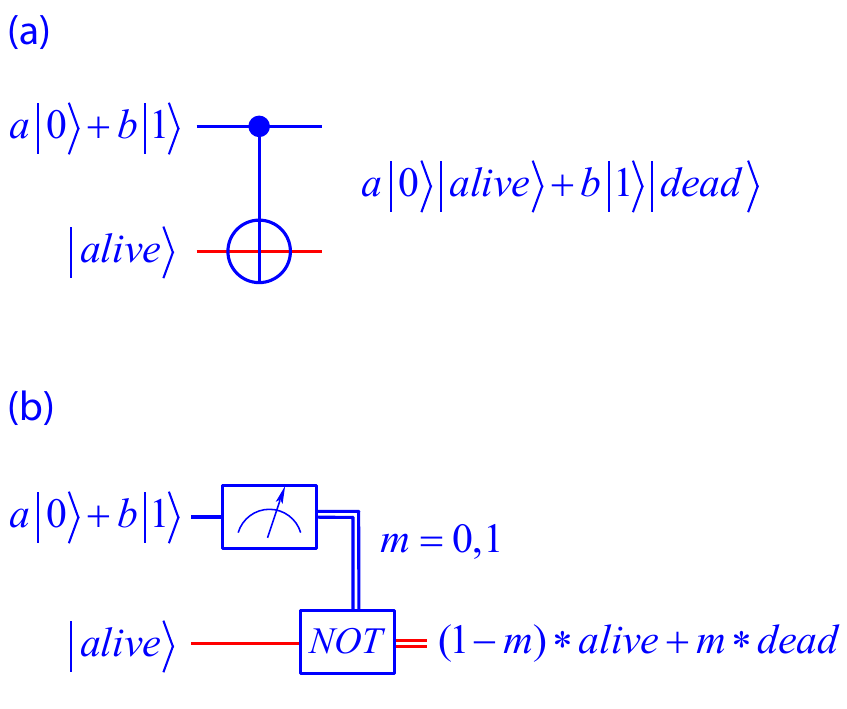}
\caption{\color[HTML]{0000FF}{\label{cat} The quantum network model reveals the information flow in the thought experiment. Single lines represent quantum systems (qubits) and double lines classical systems (bits); blue (red) lines denote the atom (cat). Quantum (classical) information flows along the single (double) lines. (a) Quantum network entangling the atom and the cat. A quantum cat in the initial state $\ket{alive}$ is coupled to the atom via a quantum CNOT gate resulting in the state $a\ket{0} \ket{alive}+ b \ket{1} \ket{dead}$. (b) Quantum network for the standard setup of Schr\"odinger's cat. The atom evolves freely to $a\ket{0}+ b\ket{1}$ and is coupled to the cat via a classical information channel: if the detector clicks ($m=1$), a poison is released and kills the cat.}}
\end{figure}

Consider first the entangled state \eqref{SC}. In order to entangle two systems, initially in a separable state, we need a quantum interaction, for example an entangling gate or a quantum channel (like in entanglement swapping). One can view this operation in the standard quantum network formalism, Fig.\ref{cat}(a). Therefore we need to apply a quantum CNOT gate between the atom and the cat in order to entangle them. The gate should be fully quantum, meaning it should operate also on quantum superpositions, not only on classical (basis) states.

However, this is not what happens in the usual setup of the thought experiment. The atom evolves freely to $a\ket{0}+ b\ket{1}$ and the only interaction between the atom and the cat is through the detector: if the detector clicks, a poison is released and the cat is killed with a \emph{classical device}, Fig.\ref{cat}(b). This is a completely classical information channel between the atom and the cat. There is no coherent quantum CNOT gate acting between the atom (the control) and the cat (the target). No amount of local operations and classical communication can increase the entanglement between two systems. In particular, two initially separable systems cannot become entangled using only local operations and classical communication. Entanglement is generated only by a quantum interaction, as can be seen in the experimental generation of cat-states \cite{Gao2010, Sychev2017}.

This implies that the final state of the atom--cat system in Fig.\ref{cat}(b) is not described by the entangled state \eqref{SC}, but by a classical statistical mixture
\begin{equation}
\rho= |a|^2 \ket{0}\bra{0} \otimes \ket{alive}\bra{alive} + |b|^2 \ket{1}\bra{1} \otimes \ket{dead}\bra{dead}
\label{rho}
\end{equation}
As expected, $\rho$ is separable since the concurrence is $C(\rho)=0$. This is in contrast to the entangled state \eqref{SC} which has $C(\ket{\psi_{\textrm f}})= 2|a \cdot b|$.

To obtain the state \eqref{SC}, which is the crux of the paradox, we need to:
\begin{enumerate}[(a)]

\item find the appropriate Hilbert space describing the cat. The cat becomes now a quantum system with at least two orthogonal states $\ket{alive}$ and $\ket{dead}$. Here we leave aside the controversial issue of {\em how to put a cat in a ket}, that is, what is the Hilbert space of a cat? what is its dimension? etc;

\item apply (experimentally) a quantum CNOT gate between the atom and the quantum cat.

\end{enumerate}

Thus, unless we are able to perform a fully coherent quantum CNOT gate between the atom and the cat, the final state of the system is described by the statistical mixture \eqref{rho}, not by the entangled state \eqref{SC}. Closing the box and refraining from looking inside (in order to prevent the collapse) will not entangle the atom and the cat. The atom still evolves freely and the only coupling between the two subsystems is via a classical channel which does not generate entanglement.

We can better understand the difference between classical and quantum information flow by looking at another thought experiment. In the quantum delayed-choice experiment, we replace the classical control in the Wheeler's delayed-choice \cite{Wheeler1984} with a quantum control \cite{Ionicioiu2011}. Although the classical and the quantum delayed-choice thought experiments are related, the quantum version gives rise to distinct phenomena which are not present in the classical case. These include a {\em morphing behaviour} between wave and particle and the ability to measure the control qubit {\em after} we measure the system \cite{Ionicioiu2011, Celeri2014}.

Interestingly, Schr\"odinger never wrote the entangled state \eqref{SC}, which later became the textbook description of the experiment \cite{Peres2002}. The following is the English translation (by J.~Trimmer) of Schr\"odinger's original article \cite{Schrodinger1935}:
\begin{quote}
One can even set up quite ridiculous cases. A cat is penned up in a steel chamber, along with the following diabolical device (which must be secured against direct interference by the cat): in a Geiger counter there is a tiny bit of radioactive substance, {\em so} small, that {\em perhaps} in the course of one hour one of the atoms decays, but also, with equal probability, perhaps none; if it happens, the counter tube discharges and through a relay releases a hammer which shatters a small flask of hydrocyanic acid. If one has left this entire system to itself for an hour, one would say that the cat still lives {\em if} meanwhile no atom has decayed. The first atomic decay would have poisoned it. The $\psi$-function of the entire system would express this by having in it the living and the dead cat (pardon the expression) mixed or smeared out in equal parts.

It is typical of these cases that an indeterminacy originally restricted to the atomic domain becomes transformed into macroscopic indeterminacy, which can then be {\em resolved} by direct observation.
\end{quote}

\section{Discussion}

The Schr\"odinger's cat paradox baffled countless physicists and laymen alike. After more than 80 years, the paradox is still unsolved and generates fierce debates about the measurement problem and the quantum-classical cut. Some of the attempts to solve the paradox \cite{Vedral2016} invoke \mbox{Everett's} many-worlds interpretation or decoherence. Nevertheless, this still does not explain how one can entangle two separable systems using only a classical channel.

The confusion behind the paradox is well-captured by Jaynes \cite{Jaynes1990, Pusey2012}:
\begin{quote}
But our present (quantum mechanical) formalism is not purely epistemological; it is a peculiar mixture describing in part realities of Nature, in part incomplete human information about Nature -- all scrambled up by Heisenberg and Bohr into an omelette that nobody has seen how to unscramble. Yet we think that the unscrambling is a prerequisite for any further advance in basic physical theory. For, if we cannot separate the subjective and objective aspects of the formalism, we cannot know what we are talking about; it is just that simple.
\end{quote}
\pagebreak
The (infamous) measurement problem is masterfully discussed in John Bell's (less famous) article entitled {\em Against `measurement'} \cite{Bell1990}. Here we did not attempt to solve the measurement problem, which would need a (presently missing) quantum ontology \cite{Ionicioiu2015}. Our goal was more modest: we showed that, in the usual description of the Schr\"odinger's cat paradox \cite{Schrodinger1935, Peres2002}, we cannot entangle the cat and the atom via a classical channel, that is, a detector. In order to entangle the two systems we need a {\em coherent quantum interaction} between the two systems.

Indeed, the experiments preparing cat-states $\tfrac{1}{\sqrt 2}(\ket{000\ldots 0}+\ket{111\ldots 1})$ use entangling quantum interactions between the subsystems \cite{Gao2010, Sychev2017}. However, such an interaction is absent in the usual description of the paradox in which a detector click starts a classical chain of events killing the cat.

In conclusion, Schr\"odinger's cat is {\em either} alive {\em or} dead, not both simultaneously. This kills the paradox (but not the cat).

\section*{Acknowledgments}

I am grateful to Daniel Terno for comments and suggestions. This work has been funded by the Romanian Ministry of Research and Innovation through grant PN~16420101/2016.

\end{document}